\documentclass[12pt]{article}

\usepackage{graphicx}            
\usepackage{psfrag}              
\usepackage{amsmath}             
\usepackage{amssymb}             
\usepackage{setspace}            
\usepackage{slashed}             
\usepackage{verbatim}            
\usepackage{array}               
\usepackage{pst-all}             

\makeatletter
\let\old@startsection=\@startsection
\renewcommand{\@startsection}[6]{\old@startsection{#1}{#2}{#3}{#4}{#5}{#6\mathversion{bold}}}
\makeatother

\newcommand{\bibtitle}[1]{{\em #1}}
\newcommand{\xxx}[1]{{\tt hep-th/#1}}

\newcommand{\ft}[2]{{\textstyle\frac{#1}{#2}}\,}
\DeclareMathOperator{\tr}{tr}                               
\DeclareMathOperator{\diag}{diag}                           
\newcommand{\be}{\begin{equation}}
\newcommand{\ee}{\end{equation}}
\newcommand{\bea}{\begin{eqnarray}}
\newcommand{\eea}{\end{eqnarray}}
\newcommand{\nn}{\nonumber}
\newcommand{\textindex}[1]{{\scriptscriptstyle\mathrm{#1}}}
\newcommand{\gym}{g_\textindex{YM}}                         
\newcommand{\Ncal}{\mathcal{N}}                             
\newcommand{\Rcal}{\mathcal{R}}                             
\newcommand{\R}{\mathbb{R}}                                 
\newcommand{\C}{\mathbb{C}}                                 
\newcommand{\unit}{\mathbf{1}}                              
\newcommand{\modulus}[1]{{| #1 |}}                          
\newcommand{\eps}{\varepsilon}                              
\newcommand{\comm}[2]{[#1,#2]}                              
\newcommand{\acomm}[2]{\{#1,#2\}}                           
\newcommand{\grU}{\mathrm{U}}                               
\newcommand{\grSU}{\mathrm{SU}}                             
\newcommand{\grSO}{\mathrm{SO}}                             
\newcommand{\ket}[1]{\bigl|#1\bigr>}                        
\newcommand{\bra}[1]{\bigl<#1\bigr|}                        
\newcommand{\braket}[2]{\bigl<#1|#2\bigr>}                  
\renewcommand{\O}{\mathcal{O}}                              
\newcommand{\state}[3]{(\mathbf{#1},\mathbf{#2},\mathbf{#3})} 

\newcommand{\x}{{\mathbf x}}                                

\newcommand{\halpha}{{\hat\alpha}}                          
\newcommand{\hbeta}{{\hat\beta}}
\newcommand{\haa}{{\hat a}}                                 
\newcommand{\hbb}{{\hat b}}
\newcommand{\hcc}{{\hat c}}

\begin{document}
\thispagestyle{empty}
\thispagestyle{empty}
\begin{flushright}
{\sc\footnotesize hep-th/0306054}\\
{\sc AEI-2003-048}
\end{flushright}
\vspace{1cm}
\setcounter{footnote}{0}
\begin{center}
{\Large{\bf \mathversion{bold} Plane-wave Matrix Theory from \par
$\Ncal=4$ Super Yang-Mills on $\R\times S^3$ \mathversion{normal}
\par}
    }\vspace{20mm}
{\sc Nakwoo Kim, Thomas Klose and Jan Plefka}\\[7mm]
{\it Max-Planck-Institut f\"ur Gravitationsphysik\\
Albert-Einstein-Institut\\
Am M\"uhlenberg 1, D-14476 Golm, Germany}\\ [2mm]
{\tt kim,thklose,plefka@aei.mpg.de}\\[20mm]

{\sc Abstract}\\[2mm]
\end{center}

\hspace*{-.4cm} Recently a mass deformation of the maximally
supersymmetric Yang-Mills quantum mechanics has been constructed
from the supermembrane action in eleven dimensional plane-wave
backgrounds. However, the origin of this plane-wave matrix theory
in terms of a compactification of a higher dimensional
Super Yang-Mills model has remained obscure.
In this paper we study the Kaluza-Klein reduction of
$D=4$, $\Ncal=4$ Super Yang-Mills theory on a round three-sphere,
and demonstrate that the plane-wave matrix theory arises through
a consistent truncation to the lowest lying modes. We further
explore the relation between the dilatation operator of the
conformal field theory and the hamiltonian of the quantum
mechanics through perturbative calculations up to two-loop order.
In particular we find that the one-loop anomalous dimensions
of pure scalar operators are completely captured by the
plane-wave matrix theory. At two-loop level this property ceases
to exist.

\newpage

\setcounter{page}{1}


\section{Introduction}

At present the most promising candidates for a microscopic description of M-theory are given in terms of
supersymmetric gauge quantum mechanical models, which are believed to provide a light-cone quantization
of M-theory in suitable backgrounds 
\cite{bfss}. Concretely in the case of a flat Minkowski background the associated matrix model is the
 maximally supersymmetric Yang-Mills quantum mechanics \cite{SYMQM}, obtained through the
discretization of the Minkowski background supermembrane action in light-cone gauge 
\cite{dWHN}. In the same spirit a similar matrix model has been proposed in \cite{bmn} for M-theory in the 
maximally supersymmetric plane-wave background first discovered by Kowalski-Glikman \cite{kowalski} in 1984,
\be
ds^2 = -2\, dx^+\, dx^- + \sum_{I=1}^9\, (dx^I)^2 - \left[ \sum_{a=1}^3\, \tfrac{\mu^2}{9}\, (x^a)^2 + \sum_{i=4}^9\, 
\tfrac{\mu^2}{36}\, (x^i)^2\, \right] \, (dx^+)^2
\ee
with non-vanishing four-form field strength $F_{123+}=\mu$. The corresponding matrix model is given by a mass 
deformation of the flat space matrix theory and takes the relatively simple form in the conventions of \cite{kimplefka}
\be \label{MMlagrangian}
 S=S_\textindex{flat} + S_M
\ee
with
\bea
S_\textindex{flat} & = & \int\!dt\: \tr \Bigl[
  \tfrac{1}{2} (D_t X_I)^2
  - i \theta D_t \theta
  + \tfrac{1}{4} \comm{X_I}{X_J}^2
  + \theta \Gamma^I \, \comm{X_I}{\theta} \, \Bigr] \; , \nn \\
S_M                   & = & \int\!dt\: \tr \Bigl[
  - \tfrac{1}{2} (\tfrac{m}{3})^2 \, (X_a)^2
  - \tfrac{1}{2} (\tfrac{m}{6})^2 \, (X_i)^2
  + \tfrac{m}{4} i\, \theta \Gamma_{123} \theta
  + \tfrac{m}{3} i \eps_{abc}\, X_a\, X_b\, X_c\, \Bigr] \; . \nn
\eea
Here $X$ and $\theta$ denote bosonic and fermionic Hermitian $N\times N$ matrices, respectively. The transverse
$\grSO(9)$ index 
$I=1,\ldots, 9$ is decomposed into $a=1,2,3$ and $i=4,\ldots ,9$ reflecting the $\grSO(3)\otimes\grSO(6)$ split of the 
transverse sector induced by the background geometry. In the above $m$ denotes the dimensionless parameter 
$m=\mu\alpha'/(2R)$, $R$ is the radius of the compactified eleventh direction and the covariant derivative is given by 
$D_t{\cal O}=\partial_t {\cal O} -i \comm{\omega}{{\cal O}}$ with the gauge field $\omega$. For $m\to 0$ ones recovers 
the usual matrix model in a flat background.

This model possesses sixteen dynamical and sixteen kinematical supersymmetries inherited from the
maximal supersymmetry of the plane-wave 
background
\bea
\delta X^I    &=& 2 \theta \Gamma^I \epsilon (t) \nn \\
\delta \theta &=& \Bigl[ i D_t X_I \Gamma^I + \tfrac{1}{2} \comm{X_I}{X_J} \Gamma^{IJ} + \tfrac{m}{3} i X_a \Gamma^a 
\Gamma_{123} - \tfrac{m}{6} i X_i \Gamma^i \Gamma_{123} \Bigr] \epsilon(t) + \eta(t) \nn \\
\delta \omega &=& 2 \theta \epsilon (t) \nn
\eea
with time-dependent parameters $\epsilon(t) = e^{-\tfrac{m}{12}\Gamma_{123}t} \epsilon_0$ and
$\eta (t) = e^{\tfrac{m}{4}\Gamma_{123}t}\eta_0$.

Plane-wave matrix theory exhibits a number of interesting properties
compared to the model without mass deformation. The usual
matrix theory has a continuous spectrum due to flat potential valleys \cite{dWLN},
whereas by virtue of the mass
terms the energy spectrum of the plane-wave matrix theory is discrete \cite{bmn,DSR1}.
Furthermore, the introduction of the dimensionless parameter $m$ allows for a
perturbative study of the spectrum of the plane-wave matrix theory for $m\gg 1$
\cite{DSR1}. As noted independently in \cite{kimplefka} and \cite{DSR2} the
spectrum contains an infinite series of protected multiplets, whose energies are
exactly given by their free field value, a property that may be
shown using the representation theory of the classical Lie
superalgebra $SU(2|4)$, the symmetry
algebra of the M-theory plane-wave, as done in \cite{DSR2} and \cite{kimpark}.
For further works on the plane-wave matrix model see 
\cite{others, park}.

Now, whereas the higher dimensional origin of the usual flat space matrix theory as the trivial  reduction of the
maximally supersymmetric $D=4$ Yang-Mills
theory to $1+0$ dimensions (or equivalently as the reduction of $\Ncal=1$ Super Yang-Mills in $1+9$ to $1+0$ 
dimensions) is obvious by taking all fields to be space-independent, a similar higher dimensional origin of the 
plane-wave matrix theory  has remained obscure.
One aim of this paper is to fill in this structural gap in the understanding of supersymmetric gauge quantum
mechanics: We shall point out that the plane-wave matrix theory arises by considering 
$\Ncal=4$, $D=4$ Super Yang-Mills compactified on a three-sphere and performing a consistent truncation of the 
resulting Kaluza-Klein spectrum. It turns out that such a truncation to a finite number of fields is only possible if one drops 
half of the vector and fermion zero modes, as we describe in detail. Upon performing this truncation we obtain 
a relation between the four-dimensional Yang-Mills coupling constant $\gym$ and the mass parameter $m$ of the 
matrix model
\be \label{mvsgym}
\left (\frac{m}{3}\right)^3 = \frac{32\pi^2}{\gym^2} \; .
\ee
Furthermore
as the radial quantization of $\Ncal=4$ Super Yang-Mills
on $\R\times S^3$ relates the energy to the dilatation operator of the
conformal field theory, one might speculate on a relation of the spectrum of the truncated model to the
scaling dimensions of composite operators in the full Super Yang-Mills field theory.
We shall show that indeed the {\sl full} one loop scaling dimensions of scalar operators in the four-dimensional 
gauge field theory are reproduced by the massive gauge quantum mechanics \eqref{MMlagrangian} in leading order 
perturbation theory. This analogy, however, breaks down at the two-loop level.

Having established the relation between the plane-wave matrix model and $\Ncal=4$, $D=4$ Super Yang-Mills, it is 
natural to seek for its possible implications on the AdS/CFT correspondence. The study of pure scalar operators is 
particularly relevant to the conjecture of Berenstein, Maldacena and Nastase (BMN) \cite{bmn} which involves operators 
whose conformal dimensions and $\grU(1)$ $R$-charges are taken to infinity. Based on holographic arguments applied to the dual 
IIB plane-wave superstring it was claimed \cite{Berenstein:2002sa} that the gauge field theory should have an effective 
one-dimensional description in the BMN limit. In fact it was proposed that this effective quantum mechanical model 
arises as a Kaluza-Klein reduction of Super Yang-Mills on $\R\times S^3$, which is precisely what we are interested 
in. The result of this paper shows that the plane-wave matrix model serves this proposal at one-loop level, 
but not at higher loops.
Put differently and slightly amusingly: The effective one-dimensional description of BMN gauge theory
is the BMN matrix model in the BMN limit -- at one-loop level and in the pure scalar sector.
Whether next-to-leading order 
corrections can be also succinctly translated into the framework of matrix quantum mechanics is a very interesting 
problem which lies beyond our scope in this paper.

The paper is organized as follows. In the next section we consider the Kaluza-Klein reduction of Super Yang-Mills on a 
three-sphere, and illustrate how it is related to the plane-wave matrix model. The effective vertices for pure 
$\grSO(6)$ bosonic excitations are constructed perturbatively in section 3 up to next-to-leading order in
perturbation theory. Our notation and some details of the computations 
are given in the appendices.


\section{$\Ncal=4$ Super Yang-Mills on $\R\times S^3$}

The field content of $D=4$, $\Ncal=4$ superconformal Yang-Mills theory consists of a vector field $A_\mu$, six real 
scalars $\phi_i$ as well as four Weyl spinors $\lambda_{\alpha A}$ all in the adjoint representation of the gauge 
group. The action of the theory on Minkowski space-time may be obtained through a trivial dimensional reduction of $\Ncal = 1$ 
supersymmetric Yang-Mills theory in $D=10$ dimensions on a six-torus. When the theory is formulated on a curved background, 
superconformal symmetry is retained if the background admits (conformal) Killing vectors and spinors which generate the 
superalgebra $\grSU(2,2|4)$. In addition there are deformations of the flat space action and the supersymmetry 
transformation rules induced by the curved background, as discussed by \cite{nicsez, okuyama} and \cite{Blau:2000xg}.
Most notably in a curved background the scalars become massive through a coupling to the Ricci scalar, reflecting the
fact that in curved spaces
the d'Alembertian operator alone is not Weyl invariant\footnote{In 
$d$-dimensions the conformally invariant wave operator is $\Box - \frac{d-2}{4(d-1)} \Rcal$, where $\Rcal$ is the Ricci 
scalar.}. In a notation where the $\grSO(1,3)\otimes\grSO(6)$ split of the $D=10$ Dirac matrices has been 
performed, the action reads
\be \label{4daction}
\begin{split}
S = \frac{2}{\gym^2} \int\!d^4x\: \sqrt{\modulus{g}} \tr \Bigl[
&
 - \tfrac{1}{4} F^{\mu\nu} F_{\mu\nu}
 - \tfrac{1}{2} D^\mu \phi_i D_\mu \phi_i
 - \tfrac{\Rcal}{12} \phi_i^2
 + \tfrac{1}{4} \comm{\phi_i}{\phi_j}^2
 -2i \lambda^{\dag}_A \sigma^\mu D_\mu \lambda^A
\\
&
+ (\rho_i)^{AB}
\lambda_{A}^\dag i \sigma^2 \comm{\phi_i}{\lambda^*_B}
 - (\rho_i^\dag)_{AB}
(\lambda^A)^\top i \sigma^2 \comm{\phi_i}{\lambda^{B}}
\Bigr] \; .
\end{split}
\ee
We use coordinates $x^\mu=(t,\theta,\psi,\chi)$ labeled by $\mu,\nu,\ldots=0,1,2,3$. When referring to spatial coordinates 
$\x^a=(\theta,\psi,\chi)$ only, we use the (curved) indices $a,b,\ldots=1,2,3$. 
The gauge covariant derivative is defined as $D_\mu = 
\nabla_\mu - i \comm{A_\mu}{\;\;}$, where $\nabla_\mu$ denotes the space-time covariant derivative. Moreover the field 
strength is given by $F_{\mu\nu} = \partial_\mu A_\nu - \partial_\nu A_\mu -i \comm{A_\mu}{A_\nu}$. We use the metric
\be
  ds^2 = g_{\mu\nu} dx^\mu dx^\nu = -dt^2 + R^2(d\theta^2 + \sin^2\theta\, d\psi^2 + \sin^2\theta\,\sin^2\psi\, 
d\chi^2)
\ee
on $\R\times S^3$, where the radius of the three-sphere $R$ is kept as a free parameter. For this background the Ricci 
scalar is $\Rcal = 6/R^2$. In \eqref{4daction} we have introduced $\sigma^\mu := (\unit, \sigma^a)$ where $\sigma^a$ 
are the usual Pauli matrices pulled back onto the $S^3$, and Clebsch-Gordan coefficients $(\rho_i)^{AB}$ of $\grSU(4)$ which relate two ${\bf 4}$'s 
with one $\bf 6$ (cf. appendix \ref{cliffordsplit} for our conventions and how the $(\rho_i)^{AB}$ are related
to the $\grSO(1,9)$ Dirac matrices).

The action \eqref{4daction} is invariant under the following modified supersymmetry transformations
\bea
\delta A_\mu    &=& 2i ( \lambda^\dag_A \sigma_\mu \eta^A - \eta^\dag_A \sigma_\mu \lambda^A ) \; , \\
\delta \phi_i   &=& -2i ( \lambda^\dag_A i \sigma^2 (\rho_i)^{AB} \eta^*_B - (\lambda^A)^\top i \sigma^2 
(\rho_i^\dag)_{AB} \eta^B) \; ,\\
\delta\lambda^A &=& \tfrac{1}{2} F_{\mu\nu} \sigma^{\mu\nu} \eta^A - D_\mu \phi_i \bar{\sigma}^\mu i\sigma^2 
(\rho_i)^{AB}\eta^*_B -\tfrac{i}{2} \comm{\phi_i}{\phi_j} (\rho_i \rho_j^\dag)^A{}_B \eta^B \nn \\
                & & -\tfrac{1}{2} \phi_i \bar{\sigma}^\mu i\sigma^2 (\rho_i)^{AB} \nabla_\mu \eta^*_B \; ,
\label{spinorsusy}
\eea
where $\bar\sigma^\mu := (-\unit,\sigma^a)$ and $\sigma^{\mu\nu} := \tfrac{1}{2} \left( \bar\sigma^\mu \sigma^\nu - 
\bar\sigma^\nu \sigma^\mu \right)$. The last term in \eqref{spinorsusy} represents the modification with respect to the flat space
transformation law. The supersymmetry parameters $\eta_{\alpha A}$ are four Weyl spinors that satisfy 
either of the two conformal Killing spinors equations:
\be \label{conformalks}
\nabla_\mu \eta = \pm \frac{i}{2R}\, \sigma_\mu \, \eta \; .
\ee
Counting the components of the Killing spinors and taking into account the degeneracy of the solutions of 
\eqref{conformalks} as well as the two signs yields the number of 32 independent real supersymmetry parameters. 
The conserved charges corresponding to $\eta_+$ and $\eta_-$ are denoted by $Q_L$ and $Q_R$ respectively. We finally 
note that the conformal Killing spinors can also be obtained from the Killing spinors of $AdS_5$ restricted on the 
boundary $\R\times S^3$.


\subsection{Harmonic expansion on $S^3$}

In order to  perform the dimensional reduction we expand the four-dimensional fields in terms of the spherical harmonics on 
$S^3$, which come in irreducible representations $(\mathbf{m_L},\mathbf{m_R})$ of the isometry group $\grSO(4) \cong 
\grSU(2)_L \otimes \grSU(2)_R$. The set of harmonics that appear in the expansion of a particular field depend on its 
spin and are listed in table~\ref{harmonics} on page \pageref{harmonics}. See also \cite{Deger:1998nm} for a useful discussion on
this topic.

\setlength{\extrarowheight}{4pt}
\begin{table}
\begin{center}
\begin{tabular}{|c|ll|c|c|} \hline
Spin & \multicolumn{2}{c|}{Harmonical functions}         &  Irreps                       & Masses              \\[2pt] 
\hline
0    & Scalar spherical harmonics:  & $Y_{(0)}^{kI}$     & $(\mathbf{k+1},\mathbf{k+1})$ & $(k+1)/R$           \\[4pt] 
\hline
1/2  & Spinor spherical harmonics:  & $Y_{(1/2)}^{kI+}$  & $(\mathbf{k+2},\mathbf{k+1})$ & $(k+\frac{3}{2})/R$ \\
     &                              & $Y_{(1/2)}^{kI-}$  & $(\mathbf{k+1},\mathbf{k+2})$ &                     \\[4pt] 
\hline
1    & Vector spherical harmonics:  & $Y_{(1)}^{kI+}$    & $(\mathbf{k+3},\mathbf{k+1})$ & $(k+2)/R$           \\
     &                              & $Y_{(1)}^{kI-}$    & $(\mathbf{k+1},\mathbf{k+3})$ &                     \\[4pt] 
\hline
\end{tabular}
\caption{\small Spherical harmonics on $S^3$ that appear in the expansion of spin-0, spin-1/2 and spin-1 fields. Here 
$k=0,1,2,\ldots$ labels different irreducible representations of $\grSU(2)$. The index $I$ enumerates the elements of a particular 
representation and therefore takes values from $1$ to the
dimension of the representation.} \label{harmonics}
\end{center}
\end{table}

We work in Coulomb gauge on $S^3$, $\nabla_a A^a = 0$, and have the following mode expansions
\begin{subequations} \label{eqn:harmonic_expansion}
\begin{align}
\phi_i(x)            & = \sum_{k=0}^\infty \sum_{I=1}^{(k+1)^2}    \phi_i^{kI}(t) Y_{(0)}^{kI}(\x) \; , \\
\lambda_\alpha^A(x)  & = \sum_{k=0}^\infty \sum_{I=1}^{(k+1)(k+2)} \sum_\pm \lambda^{A,kI\pm}(t) Y_{(1/2)\:\alpha
}^{kI\pm}(\x) \; , \\
A_0(x)               & = \sum_{k=0}^\infty \sum_{I=1}^{(k+1)^2}    \omega^{kI}(t) Y_{(0)}^{kI}(\x) \; , \\
A_a(x)               & = \sum_{k=0}^\infty \sum_{I=1}^{(k+1)(k+3)} \sum_\pm A^{kI\pm}(t) Y_{(1)\:a}^{kI\pm}(\x) \; .
\end{align}
\end{subequations}
Note that the spinor spherical harmonics $Y_{(1/2)\:\alpha}^{kI\pm}$ are two-dimensional commuting Weyl spinors.
Upon inserting the above harmonic expansions into the action \eqref{4daction} and carrying out the integration over the
three-sphere one obtains a one-dimensional theory with an infinite number of fields. In order to determine the mass
spectrum  of these excitations we shall perform this integration for
the quadratic terms of the action \eqref{4daction}. For this we need
only a few properties of the spherical harmonics. They are orthonormalized to
\be
\int_{S^3}  Y_{(0)}^{kI} \; Y_{(0)}^{lJ} = \delta^{kl} \delta^{IJ}
     \; ,
\int_{S^3}  \bigl( Y_{(1/2)\:\alpha}^{kI\pm} \bigr)^* \; Y_{(1/2)\:\alpha}^{lJ\pm} = \delta^{kl} \delta^{IJ}
    \; ,
\int_{S^3}  Y_{(1)\:a}^{kI\pm} \; Y_{(1)}^{lJ\pm\:a} = \delta^{kl} \delta^{IJ}
     \label{eqn:harmonics_orthonormality} 
\ee
and their eigenvalues of the Laplace-Beltrami operators are
\begin{subequations} \label{scalar_harmonics_laplace}
\begin{align}
\nabla^2 Y^{kI}_{(0)}              &= -\tfrac{1}{R^2} k(k+2) Y^{kI}_{(0)} \; , \\
\slashed{\nabla} Y^{kI\pm}_{(1/2)} &= \pm \tfrac{i}{R} (k+\tfrac{3}{2}) Y^{kI\pm}_{(1/2)} \; , \\
\nabla^2 Y^{kI\pm}_{(1/2)\:\alpha}         &= -\tfrac{1}{R^2} \left[ k(k+3)+ \tfrac{3}{4} \right] Y_{(1/2)\:\alpha}^{kI\pm} \; ,\\
\nabla^2 Y^{kI\pm}_{(1)\:a}           &= -\tfrac{1}{R^2} \left[ k(k+4) + 2 \right] Y^{kI\pm}_{(1)\:a} \; .
\end{align}
\end{subequations}
The above operators include, of course, only the spatial parts $\slashed{\nabla} := \sigma^a\nabla_a$ and $\nabla^2 
:= \nabla^a \nabla_a$.

Upon inserting the mode expansions \eqref{eqn:harmonic_expansion} into the quadratic part of the action \eqref{4daction},
one obtains
\be \label{eqn:reduced_action_bilinear_part}
\begin{split}
S_{\textindex{quadratic}} = \frac{4 \pi^2 R^3}{\gym^2} \int\!dt\: \biggl[
&  \sum_{k,I}\,\biggl\{ \frac{1}{2}
\tr \dot \phi_i^{kI} \dot \phi_i^{kI} - \frac{(k+1)^2}{2R^2} \tr \phi_i^{kI} \phi_i^{kI}\biggr\}  \\
+& \sum_{k,I,\pm} \biggl\{\frac{1}{2} \tr \dot A^{kI\pm} \dot A^{kI\pm} - \frac{(k+2)^2}{2 R^2} \tr A^{kI\pm} A^{kI\pm} 
\biggr\} \\
-i &  \sum_{k,I,\pm} \biggl\{  \tr {\lambda^{kI\pm}_A}^\dag \dot \lambda^{A, kI\pm} +
 \frac{k+\frac 32}{R} \tr \lambda^{kI\pm\:\dagger}_A \, \lambda^{A, kI\pm} \biggr\}\, \biggr]
\end{split}
\ee
where we have made use of the orthonormality conditions, integration by parts and the properties 
\eqref{scalar_harmonics_laplace}. In the computation of the masses for the 
vector modes one also needs  the transversality of the vector spherical harmonics, $\nabla^a Y_{(1)\:a}^{kJ\pm}=0$,
and the  identity
\be
\nabla_a \nabla_b Y_{(1)}^{lJ\pm\:a}
= \comm{\nabla_a}{\nabla_b} Y_{(1)}^{lJ\pm\:a}
= R^a{}_{cab} Y_{(1)}^{lJ\pm\:c}
= \Rcal_{ab} Y_{(1)}^{lJ\pm\:a}
= \frac{2}{R^2} Y_{(1)\:b}^{lJ\pm}
\ee
using $\Rcal_{ab} = \frac{2}{R^2} g_{ab}$ of $S^3$.
The obtained mass spectrum is summarized in table~\ref{harmonics}.

\begin{figure}[t]
\begin{center}
\psfrag{A1}[c][c]{$A^{kI+}$} \psfrag{L1}[c][c]{$\lambda^{kI+}$}
\psfrag{PHI}[c][c]{$\phi^{kI}$} \psfrag{L2}[c][c]{$\lambda^{kI-}$}
\psfrag{A2}[c][c]{$A^{kI-}$}
\psfrag{st421}[c][c]{$\state{4}{2}{1}$}
\psfrag{st336}[c][c]{$\state{3}{3}{6}$}
\psfrag{st241}[c][c]{$\state{2}{4}{1}$}
\psfrag{st324}[c][c]{$\state{3}{2}{4}$}
\psfrag{st234}[c][c]{$\state{2}{3}{\bar 4}$}
\psfrag{sta}[c][c]{$\state{3}{1}{1}$}
\psfrag{st226}[c][c]{$\state{2}{2}{6}$}
\psfrag{st131}[c][c]{$\state{1}{3}{1}$}
\psfrag{stlambda}[c][c]{$\state{2}{1}{4}$}
\psfrag{st124}[c][c]{$\state{1}{2}{\bar 4}$}
\psfrag{stphi}[c][c]{$\state{1}{1}{6}$} \psfrag{QL}[c][c]{$Q_L$}
\psfrag{QR}[c][c]{$Q_R$} \psfrag{mass}[c][c]{Mass}
\psfrag{mass4}[c][c]{$\tfrac{3}{R}$}
\psfrag{mass3}[c][c]{$\tfrac{5}{2R}$}
\psfrag{mass2}[c][c]{$\tfrac{2}{R}$}
\psfrag{mass1}[c][c]{$\tfrac{3}{2R}$}
\psfrag{mass0}[c][c]{$\tfrac{1}{R}$}
\includegraphics*{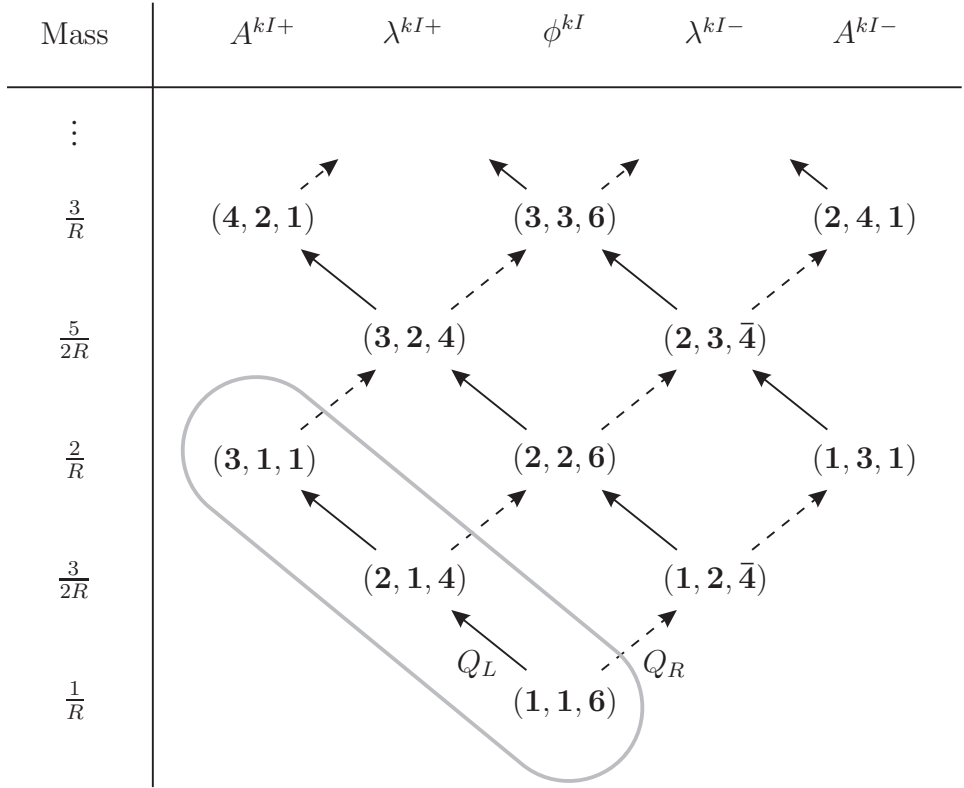}
\end{center}
\caption{\small Kaluza-Klein particle spectrum of $\Ncal=4$ Super Yang-Mills on $\R\times S^3$. The states are labeled 
by representations of $\grSU(2)_L \otimes \grSU(2)_R \otimes \grSU(4)$. One climbs upward to the left by acting with 
the supercharge $Q_L=\state{2}{1}{\bar 4}$ and upward to the right with $Q_R=\state{1}{2}{4}$. The encircled states
comprise the consistent truncation to the plane-wave matrix theory.}
\label{tab:particle_spectrum}
\end{figure}

In figure~\ref{tab:particle_spectrum} on page~\pageref{tab:particle_spectrum} we present the resulting Kaluza-Klein 
mass tower up to mass $\frac 3 R$. One may climb up the various states of the tower by acting with the two supercharges 
$Q_L=\state{2}{1}{\bar 4}$ (to the upper-left) and $Q_R=\state{1}{2}{4}$ (to the upper-right). Note that unlike the 
trivial dimensional truncation on flat spaces, where superpartners have the same masses,
the superconformal transformations in our curved background geometry relate the entire tower of
Kaluza-Klein modes. In other words we 
find that the infinite tower of Kaluza-Klein modes is not decomposed into finite dimensional irreducible 
representations of the superconformal algebra. Instead, the entire tower itself is a single irreducible representation.

On the other hand, if we consider only half of the supercharges, say $Q_L$, which together with the bosonic symmetries
generate the subalgebra $\grSU(2|4)$, the Kaluza-Klein modes are decomposed in terms of finite dimensional irreducible 
representations. We see that $\bf (1,1,6) + (2,1,4) + (3,1,1)$ (encircled in figure 1)
is the lowest such supermultiplet. Higher ones in 
general branch into five irreducible representations under the bosonic subalgebra $\grSU(2) \otimes \grSU(4)$, implying 
that they are all short supermultiplets of $\grSU(2|4)$. In particular the lowest lying multiplet consisting of three
floors is $1/2$ BPS and the following multiplet comprised of four floors is $1/4$ BPS, as may be deduced from the
results of \cite{kimpark}. An important result of the supersymmetry algebra is that the 
ground state energy given as the sum of zero point energies should vanish, which we can easily verify. For instance, 
for the lowest lying supermultiplet $\bf (1,1,6) + (2,1,4) + (3,1,1)$ the summation goes as follows
\be
6\cdot \frac{1}{R} - 8\cdot \frac{3}{2R} + 3\cdot \frac{2}{R} = 0 \; .
\ee
The truncation we are about to perform consists precisely in the restriction to this lowest lying
multiplet.

\subsection{Derivation of Plane-wave matrix theory}\label{mm_derivation}

We now aim to show that the infinite tower of Kaluza-Klein states can be consistently truncated to the lowest lying 
supermultiplet, and that the one-dimensional action, which governs its dynamics, is precisely the plane-wave matrix 
model. Hence we restrict the expansion \eqref{eqn:harmonic_expansion} to the zero modes which are $\grSU(2)_R$ 
singlets. For these modes the harmonic functions are given by a constant, two Killing spinors $S^{\halpha+}_\alpha$, and three 
Killing vectors $V^{\haa+}_a$. The two Killing spinors obey the defining equation
\be
\nabla_a S^{\halpha+}_\alpha= \frac{i}{2R}\, (\sigma_a)_{\alpha\beta}\, S^{\halpha+}_\beta, \qquad \halpha=1,2
\label{KSeq}
\ee
and lead to the three positive chirality Killing vectors $V^{\haa+}_a$ via\footnote{The
Killing vectors of opposite chirality $V^{\haa-}_a$ are induced from the opposite chirality Killing spinor
bilinear $S^{\halpha-\:\dagger} \sigma_a S^{\hbeta-}$ which obey \eqref{KSeq} with an opposite sign.}
\be
S^{\halpha+\:\dagger} \sigma_a S^{\hbeta+}= (\sigma_\haa)^{\halpha\hbeta}\, V_a^{\haa+}\; .
\ee
Their explicit form along with further useful properties may be found in appendix \ref{killing}.
The truncated mode expansion then takes the form
\begin{subequations} \label{eqn:ansatz_zero_modes}
\begin{align}
  \phi_i(x)           & = X_i (t) \; , \\
  \lambda^A_\alpha(x) & = \sum_{\halpha=1}^2 \theta^A_\halpha (t) S_\alpha^{\halpha+}(\x) \; , \\
  A_0(x)              & = \omega (t) \; , \\
  A_a(x)              & = \sum_{\haa=1}^3 X_{\haa} (t) V_a^{\haa+}(\x) \; .
\end{align}
\end{subequations}

In order to prove that the above ansatz is a consistent truncation of the spectrum, we have to insert this ansatz into 
the four-dimensional equations of motion and show that this leads to a set of equations without contradictions. The 
equations of motions which follow from the four-dimensional action \eqref{4daction} are given by
\begin{subequations} \label{eqn:4d_eom}
\begin{align}
& D^\nu F_{\nu\mu} + i \comm{\phi_i}{D_\mu \phi_i}
  + 2 \acomm{ \lambda^\dag }{ \sigma_\mu \lambda }
  = 0 \; , \label{eqn:eom_vector} \\
& D^\mu D_\mu \phi_i
  - \tfrac{1}{R^2} \phi_i
  + \comm{ \phi_j }{ \comm{ \phi_i }{ \phi_j } }
  - \acomm{ \lambda^\dag }{ i\sigma^2 \rho_i \lambda^* }
  + \acomm{ \lambda^\top }{ i \sigma^2 \rho_i^\dag \lambda }
  = 0 \; , \label{eqn:eom_scalar} \\
& i \sigma^\mu D_\mu \lambda
  - i \sigma^2 \rho_i \comm{\phi_i}{\lambda^*}
  = 0 \; . \label{eqn:eom_spinor}
\end{align}
\end{subequations}
where the anticommutator of two fermions is defined as $\acomm{\psi^\top}{\chi}^A := if^{ABC} (\psi^\top)^B \chi^C$. 
Inserting the ansatz \eqref{eqn:ansatz_zero_modes}
is a straight-forward computation which we comment on in appendix \ref{derivation_eom}. One finds that 
the $S^3$ coordinate dependences of all terms in each equation exactly matches, and that the $D=4$
equations of motion are satisfied provided that the one-dimensional fields $X^I(t), \omega(t)$ and $\theta(t)$ obey
\begin{subequations} \label{eqn:mm_eom}
\begin{align}
& \comm{X_I}{i D_t X_I}
  - 2 \acomm{ \theta^{\dag} }{ \theta }
  = 0 \; , \label{eqn:mm_gauge_condition} \\
& D^2_t X_{\haa}
  + \tfrac{4}{R^2} X_{\haa}
  - \tfrac{6i}{R} \eps_{\haa\hbb\hcc} X_{\hbb}X_{\hcc}
  - \comm{ X_I }{ \comm{ X_{\haa} }{ X_I } }
  - 2 \acomm{ \theta^{\dag} }{ \sigma_{\haa} \theta }
  = 0 \; , \label{eqn:mm_eom_so3_scalars} \\
& D^2_t X_i
  + \tfrac{1}{R^2} X_i
  - \comm{ X_I }{ \comm{ X_i }{ X_I } }
  + \acomm{ \theta^{\dag} }{ i \sigma^2 \rho_i \theta^{*} }
  - \acomm{ \theta^\top }{ i \sigma_2 \rho_i^\dag \theta }
  = 0 \; , \label{eqn:mm_eom_so6_scalars} \\
& i D_t \theta
  - \tfrac{3}{2R} \theta
  + \comm{ X_{\haa} }{ \sigma_{\haa} \theta }
  - \comm{ X_i }{ i \sigma_2 \rho_i \theta^* }
  = 0 \; . \label{eqn:mm_eom_spinors}
\end{align}
\end{subequations}
Note that the label $\halpha$ of the fermion zero modes $\theta^A_\halpha$ which used to account for their 
degeneracy has turned into a proper (Weyl) spinor index. Having found the equations of motion \eqref{eqn:mm_eom} for 
the zero modes, one may try to find a quantum mechanical action which leads to these equations of motion.
It can be checked easily that these 
equations are derived from the following one-dimensional Lagrangian
\be
\begin{split}
L =
& \tr \Bigl[
\tfrac{1}{2} (D_t X_I)^2 -\tfrac{1}{2} (\tfrac{m}{3})^2 X_a^2
-\tfrac{1}{2} (\tfrac{m}{6})^2 X^2_i
+\tfrac{m}{3} i \eps_{abc} X_a X_b X_c
+\tfrac{1}{4} \comm{X_I}{X_J} ^2
\\
&
- 2 i \theta^\dag D_t \theta +\tfrac{m}{2} \theta^\dag \theta
- 2 \theta^\dag \sigma^a \comm{X_a}{\theta}
+ \theta^\dag i\sigma^2 \rho_i \comm{X_i}{\theta^*}
- \theta^\top i \sigma^2 \rho^\dag_i \comm{X_i}{\theta}
\Bigr]
\end{split}
\ee
where the radius of the three-sphere $R$ has been traded for a mass parameter $m=6/R$. This lagrangian is in fact 
nothing but the plane-wave matrix theory \eqref{MMlagrangian} written in terms of $\grSU(2)\otimes\grSU(4)$
spinor variables. This dimensional split was performed for instance in \cite{park},
in appendix \ref{cliffordsplit} we repeat this split using our 
conventions and notation for easier reference.

This result also induces a relation of the four-dimensional Yang-Mills coupling constant
$\gym$ to the mass parameter $m$ of the plane-wave matrix model. It is found by requiring the prefactor
of the reduced Super Yang-Mills action \eqref{eqn:reduced_action_bilinear_part} to match the unit prefactor
of the matrix model action \eqref{MMlagrangian}. Taking into account that $m=6/R$ one finds
\be
  \left( \frac{m}{3} \right)^3 = \frac{32 \pi^2}{\gym^2} \: .
\ee


\section{Perturbation Theory}

Having established this connection of $D=4$, $\Ncal=4$ Super Yang-Mills to the plane-wave matrix model one might wonder what 
additional structures of the field theory carry over to the matrix quantum mechanics. By virtue of the state-operator 
map of conformal field theories one might expect a connection of scaling dimensions of Super Yang-Mills operators on 
$\R^4$ to the energy of corresponding states in the plane-wave matrix model. As noted in \cite{kimplefka,DSR2} one such 
connection is already manifest: The  protected multiplets of chiral primary operators in the gauge theory, their 
lightest representative being the pure multi-scalar operators $\O_{CPO}=c_{i_1\ldots i_n}\,\tr 
(\phi_{i_1}\ldots\phi_{i_n})$, with $c_{i_1\ldots i_n}$ symmetric and traceless, have vanishing anomalous dimensions, 
i.e.~their scaling dimension is exactly given by their free field value $\Delta_{\O_{CPO}}=n$. This property is 
paralleled by the existence of protected multiplets in the plane-wave matrix theory, with their lightest member again being 
the symmetric traceless excitations $c_{i_1\ldots i_n}\, \tr(a^\dag_{i_1}\ldots a^\dag_{i_n})|0\rangle$ in the 
$\grSO(6)$ sector of the matrix model \eqref{MMlagrangian} \cite{kimplefka,DSR2}. The energy of these states is 
similarly not corrected by interactions and simply given by the free theory value $E=n\cdot\tfrac{m}{6}$ to all orders 
in perturbation theory\footnote{Note that only the energy eigenvalue is protected, the eigenstates are renormalized in 
perturbation theory.}. So here a complete analogy exists and it is natural to ask whether it extends to non-protected 
states in the $\grSO(6)$ sector as well. That is, is there a general relation between the scaling dimension in 4d and 
the quantum mechanical energy? As we will show in the following this analogy extends to the one loop level in the 
$\grSO(6)$ sector, but breaks down at two loops. Technically what we demonstrate is the precise matching of the 
effective quantum mechanical interaction vertex in first order perturbation theory with the one loop piece of the 
dilatation operator of $D=4$, $\Ncal=4$ Super Yang-Mills computed in \cite{Boston,Potsdam}. Comparing our second-order 
perturbation theory result to the recently established two-loop piece of the dilatation operator \cite{Beisert:2003tq} 
we find a discrepancy, although the same structure of terms does appear. One might still hope that this discrepancy 
disappears once one considers the BMN limit of the gauge theory, however, this turns out to not be the case.

Let us then reconsider the perturbative evaluation of energy shifts in the $\grSU(N)$ plane-wave matrix model discussed in 
\cite{DSR1,kimplefka}. The hamiltonian associated to \eqref{MMlagrangian} may be split into a free and an interaction 
piece. Working in the conventions of \cite{kimplefka} the free piece reads ($a=1,2,3$; $i=1,\ldots,6$; 
$\alpha=1,\ldots,16$)
\be
  H_0 :=\tr \left[ \frac{m}{6} a_i^\dag a_i +
\frac{m}{3} a_a^\dag a_{a} + \frac{m}{2} \theta_\alpha^+
\theta^-_\alpha \right]
\ee
with the matrix oscillators
\begin{align}
a_i        & := \sqrt{\tfrac{3}{m}} \left( P_i - i \tfrac{m}{6} X_i \right)
& \theta^\pm_\alpha     & := \Pi_{\alpha\beta}\, \theta_\beta \nn \\
a_a     & := \sqrt{\tfrac{3}{2m}}  \left( P_a - i \tfrac{m}{3} X_a \right)
& \Pi^\pm_{\alpha\beta} & :=\tfrac{1}{2}(\unit\pm i \Gamma_{123})_{\alpha\beta}
\end{align}
obeying
\begin{align}
  \comm{(a_i)_{rs}}{(a_j^\dag)_{tu}}       & =\delta_{ij}       \left( \delta_{st} \delta_{ru} - \tfrac{1}{N} 
\delta_{rs} \delta_{tu} \right) \nn\\
  \comm{(a_a)_{rs}}{(a_b^\dag)_{tu}} & =\delta_{ab} \left( \delta_{st} \delta_{ru} - \tfrac{1}{N} 
\delta_{rs} \delta_{tu} \right) \\
  \acomm{(\theta_\alpha^-)_{rs}}{(\theta_\alpha^+)_{tu}}
                                           & = \tfrac{1}{2} (\Pi^-)_{\alpha\beta} \left( \delta_{st} \delta_{ru} - 
\tfrac{1}{N} \delta_{rs} \delta_{tu} \right)\, . \nn
\end{align}
The interacting piece of the hamiltonian is comprised of cubic and quartic terms in the fields
\be
H_{\rm int} := V_1 + V_2
\ee
where
\begin{align}
V_1 & = - \frac{m}{3} i\eps_{abc} \tr X_a X_b X_c - \tr \theta \gamma_I \comm{X_I}{\theta} \nn \\
V_2 & = - \frac{1}{4} \tr \comm{X_I}{X_J} \comm{X_I}{X_J} \nn
\end{align}
and $I$ is a joint index $I=(a,i)$. For $m\gg 1$ the interaction piece $H_{\rm int}$ may be treated 
perturbatively\footnote{This is easily seen by performing a rescaling $X \to X/m$ and $\theta \to \theta/\sqrt{m}$, 
rendering the hamiltonian into the form $H = \tilde H_0 + \tfrac{1}{m^2} \tilde V_1 + \tfrac{1}{m^4} \tilde V_2$ with 
no $m$ dependence in $\tilde H_0$, $\tilde V_1$ and $\tilde V_2$.}. The first order energy shift of a state 
$\ket{\phi_0}$ reads
\be
  \delta E_1 = \bra{\phi_0} V_{\rm eff}^{(1)} \ket{\phi_0}
\qquad
\mbox{with}
\quad
  V_{\rm eff}^{(1)} = V_1 \Delta V_1 + V_2
\qquad
\mbox{and}
\quad
\Delta = \frac{\unit - \ket{\phi_0}\bra{\phi_0}}{E_0 - H_0}
\ee
where we assume $\braket{\phi_0}{\phi_0} = 1$ and a non-degenerate free energy $E_0$. As we shall only investigate the 
energy shifts of pure $\grSO(6)$ excitations we will normal order $V_{\rm eff}^{(1)}$ and drop all normal ordered terms 
containing fermions or $\grSO(3)$ oscillators, as these would annihilate a pure $\grSO(6)$ excitation state. We 
indicate this simplification by an arrow. Using {\tt Mathematica} and {\tt Form} \cite{Jos} we find
\begin{align}
V_1 \Delta V_1 \rightarrow\ & - \frac{99}{4M^2} (N^3 - N)
- \frac{12N}{M^2} : \tr a_i^\dag a_i : \\
V_2                \rightarrow\ & \frac{99}{4M^2} (N^3 - N)
                                + \frac{13N}{M^2} : \tr a_i^\dag a_i :
                                + \frac{1}{2M^2} : \tr \comm{a_i^\dag}{a_i}
\comm{a_j^\dag}{a_j} : \nonumber \\
       & - \frac{1}{2M^2} : \tr \comm{a_i^\dag}{a_j} \comm{a_i^\dag}{a_j} :
               - \frac{1}{M^2}: \tr \comm{a_i^\dag}{a_j^\dag} \comm{a_i}{a_j} :
\end{align}
introducing the shorthand $M:=m/3$ which controls the perturbative expansion. When adding these contributions all 
constants sum to zero
\be
V_{\rm eff}^{(1)} \rightarrow
\frac{1}{M^2}\biggl(
N : \tr a_i^\dag a_i : + \frac{1}{2} : \tr \comm{a_i^\dag}{a_i}
\comm{a_j^\dag}{a_j} : - \frac{1}{2} : \tr \comm{a_i^\dag}{a_j}
\comm{a_i^\dag}{a_j} : - : \tr
\comm{a_i^\dag}{a_j^\dag} \comm{a_i}{a_j} :   \biggr)
\ee
a manifestation of the vanishing energy shift of the groundstate $\ket{0}$. This expression can be simplified further 
as the state $\ket{\phi_0}$ acting upon it is gauge invariant, e.g.~some multi-trace excitation. The simplification 
concerns all terms with a commutator of contracted creation and annihilation operators: $\comm{a_i^\dag}{a_i}$, 
i.e.~the second term in the above. One splits off this commutator from the rest and removes the normal ordering between 
the two factors:
\be
\label{here}
\begin{split}
\frac{1}{2} : \tr \comm{a_i^\dag}{a_i} \comm{a_j^\dag}{a_j} :
   & = \frac{1}{2} : \tr \comm{a_i^\dag}{a_i} T^A \tr T^A \comm{a_j^\dag}{a_j} : \\
   & = \frac{1}{2} \left( : \tr \comm{a_i^\dag}{a_i} T^A : \right) \left( : \tr T^A \comm{a_j^\dag}{a_j} : \right)
     - N : \tr a_i^\dag a_i :
\end{split}
\ee
where $T^A$ denote the generators of $\grSU(N)$ and we note that
\begin{align}
  \tr (T^A A T^A B)        & = \tr A \tr B - \tfrac{1}{N} \tr (AB)  \; , \\
  \tr (T^A A ) \tr (T^A B) & = \tr (A B) - \tfrac{1}{N} \tr A \tr B \; .
\end{align}
The first term in \eqref{here} can be dropped for reasons given in the
appendix \ref{sec:misc} and the second term cancels the two point interaction that was already present. Hence we arrive 
at the
compact result
\be \label{Veff1final}
V_{\rm eff}^{(1)} \rightarrow \frac{1}{M^2}\biggl(
- \frac{1}{2} : \tr \comm{a_i^\dag}{a_j} \comm{a_i^\dag}{a_j} : - : \tr \comm{a_i^\dag}{a_j^\dag} \comm{a_i}{a_j} : 
\biggr )
\ee

We want to compare this 1-loop effective vertex, which determines the first
order energy shifts of states in the plane-wave matrix model, to the 1-loop dilatation operator of 
$\Ncal=4$ Super Yang-Mills,
which determines the anomalous dimensions of operators. We take the dilatation operator from \cite{Beisert:2003tq}.
The 1-loop contribution is given in Eq. (1.6):
\be\label{D2}
D_2 = \frac{\gym^2}{16\pi^2}\left ( -\frac{1}{2} : \tr \comm{\phi_i}{\check\phi_j} \comm{\phi_i}{\check\phi_j} :
- : \tr \comm{\phi_i}{\phi_j} \comm{\check\phi_i}{\check\phi_j} : 
\right )\; ,
\ee
where $\check\phi_i:=\frac{d}{d\phi_i}$.
This shows that by identifying the scalars of both theories
\be
  a_i^\dag \hat= \phi_i \qquad , \qquad  a_i \hat= \check\phi_i \; ,
\ee
we have an exact agreement between energy shifts and anomalous dimensions for all
pure $\grSO(6)$ states/operators at one loop level!

This result is intriguing in view  of the fact \cite{Minahan:2002ve}
that the 1-loop dilatation operator $D_2$ of $\Ncal=4$ Super Yang-Mills
in the strict $N\to\infty$ limit may be viewed
as the hamiltonian of an integrable $\grSO(6)$ spin chain model,
hinting at the integrability of $\Ncal=4$
Super Yang-Mills. This structure thus carries over to the plane-wave matrix model.
The higher-loop corrections to the Super Yang-Mills dilatation operator represent
non-standard deformations of this spin chain model as discussed in
\cite{Beisert:2003tq}. The fact that the 2-loop effective vertex in
the plane-wave matrix quantum mechanics to be discussed below departs from the
Super Yang-Mills dilatation operator indicates that one is facing an alternative
deformation of the spin chain in the plane-wave matrix theory, iff the integrability is
indeed conserved at higher loop level.  Including the fermionic and $\grSO(3)$ excitations
one might expect to uncover an integrable  spin chain structure of plane-wave matrix theory
based on the underlying supergroup $\grSU(2|4)$. It is certainly interesting
to pursue these issues further.

Moving on to the 2-loop energy shift we are faced with the evaluation of the
effective vertex $V_{\rm eff}^{(2)}$
\be
  \delta E_2 = \bra{\phi_0} V_{\rm eff}^{(2)} \ket{\phi_0}
\ee
with
\be\label{Veff2}
\begin{split}
V_{\rm eff}^{(2)} =\ & V_1 \Delta V_1 \Delta V_1 \Delta V_1 \\
                     & + V_1 \Delta V_1 \Delta V_2 + V_1 \Delta V_2 \Delta V_1 + V_2 \Delta V_1 \Delta V_1 \\
                     & + V_2 \Delta V_2 - V_1 \Delta \Delta V_1 P V_1 \Delta V_1 - V_1 \Delta \Delta V_1 P V_2
\end{split}
\ee
where
\be
\Delta = \frac{\unit - \ket{\phi_0}\bra{\phi_0}}{E_0 - H_0} \; , \qquad\qquad
P      = \ket{\phi_0}\bra{\phi_0}\, .
\nn\ee
Here some information on the particular forms of $V_1$ and $V_2$ entered
in the derivation of \eqref{Veff2}: All terms have been omitted that contain an odd number of fields of one kind, as 
they do not contribute
in the computation of expectation values. We again project onto pure, gauge
invariant $\grSO(6)$ excitations acting on $V_{\rm eff}^{(2)}$. The
resulting expression is, however, still rather lengthy and has been relegated
to appendix \ref{sec:perturbation_theory_2-loop} eq.~\eqref{dasletzte}. Let us then exemplify our
claim of the disagreement at 2-loop level by considering the energy shifts
of an explicit state.

Consider the ``Konishi'' state in the plane-wave matrix theory
\be
  \ket{K} := \frac{1}{\sqrt{12(N^2-1)}} \tr a_i^\dag a_i^\dag \ket{0}\, .
\ee
We then find
\begin{align}
  E_0        & = \bra{K} H_0 \ket{K} = M \\
  \delta E_1 & = \bra{K} V_{\rm eff}^{(1)} \ket{K} = \frac{12N}{M^2} \\
  \delta E_2 & = \bra{K} V_{\rm eff}^{(2)} \ket{K} = - \frac{228N^2}{M^5}
\end{align}
This is to be compared to the scaling dimension of the Konishi operator
$K=\tr (\phi_i\phi_i)$ of $\Ncal=4$ Super Yang-Mills \cite{KonishiAD}
\be\label{konsd}
  \Delta_K = 2 + \frac{3\gym^2 N}{4\pi^2} - \frac{3\gym^4 N^2}{16\pi^4} + \ldots
\ee
As the overall constant of the hamiltonian is not fixed we
can only compare the ratios of the energy shifts $\delta E_i$ to the free energy
$E_0$ with the ratios of the anomalous dimensions $\delta \Delta_i$ to
the engineering dimension $\Delta_0$.
\be
E_{|K\rangle}=\frac M 2\, \Bigl ( 2+ \frac{24\, N}{M^3} - \frac{456\, N^2}{M^6}+\ldots\Bigr )
\ee
Inserting the uncovered relation between the plane-wave matrix theory mass parameter
$M=\frac{m}{3}$ and the gauge theory coupling constant $\gym$ of \eqref{mvsgym}
into \eqref{konsd}
\be \label{mvsgym2}
  \frac{1}{M^3} = \frac{\gym^2}{32\pi^2}
\ee
the 1-loop correction indeed agrees\footnote{An alert reader might think
that in choosing the relation \eqref{mvsgym2} one does not have an independent
check of the 1-loop agreement. This is of course true. The point here is that
after choosing \eqref{mvsgym2} {\sl all} further 1-loop corrections for higher
excited states are fixed and do agree with the corresponding
1-loop Super Yang-Mills scaling dimensions.
This is a consequence of the matching of \eqref{Veff1final} with \eqref{D2}.}.
Using this relation in the 2-loop energy shift we would predict the following
2-loop anomalous dimension for the Konishi field from the matrix quantum mechanics
\be
  \frac{19}{8} \cdot \frac{-3\gym^4 N^2}{16\pi^4}
\ee
which is off by a factor of $19/8$.

\subsection{The Berenstein-Maldacena-Nastase limit}

Recently there has been considerable interest in a novel double scaling limit
of $\grSU(N)$ $D=4$, $\Ncal=4$ Super Yang-Mills following the work of Berenstein, Maldacena
and Nastase \cite{bmn} where one takes 
\be
N\to \infty, \quad J\to \infty\quad \mbox{with} \quad \frac{J^2}{N} \quad \mbox{and}\quad \gym \quad\mbox{fixed.}
\label{BMNlimit}
\ee
Here $J$ is the $\grU(1)$ $R$-charge associated to the complex combination of two of
the six scalar fields $\phi_i$, e.g.
\be
Z=\frac{1}{\sqrt{2}}\,( \phi_5+i\,\phi_6) \; .
\ee
This limit leads to a dual gauge theory description of the IIB plane-wave superstring. It represents
an extreme reduction of the field theory: Only two- and three-point functions of so-called
BMN operators, having large scaling dimensions and $\grU(1)$ $R$-charges $\Delta_0>J\gg 1$, exist
\cite{Potsdam}. It also appears at this stage, that the only sensible observable of BMN gauge
theory is the dilatation operator. Moreover, holographic arguments indicate that the gauge 
theory dual of the IIB plane-wave superstring should be given by a one-dimensional model
and it was proposed \cite{Berenstein:2002sa} that this effective quantum mechanical model 
indeed arises as the Kaluza-Klein reduction of $D=4$, $\Ncal=4$ Super Yang-Mills on $\R\times S^3$,
which we have considered in section 2. As long as one restricts one's attention to the
pure scalar $\grSO(6)$ sector we have seen that the plane-wave matrix theory serves this
purpose at the one-loop level. In principle there is the logical possibility that although
this match of energies and scaling dimensions ceases to exist at higher loop level 
for the full Super Yang-Mills theory,
it does hold for the BMN sector of the gauge theory at higher loops.

For convenience we further reduce the $\grSO(6)$ excitations under consideration
in the plane-wave matrix model, to solely two complex combinations
\begin{align}
  Z      & := \tfrac{1}{\sqrt{2}} \left( a_5^\dag + i a_6^\dag \right)
& \bar Z      & := \tfrac{1}{\sqrt{2}} \left( a_5      - i a_6      \right) \\ 
  \phi    & := \tfrac{1}{\sqrt{2}} \left( a_3^\dag + i a_4^\dag \right)
& \bar\phi    & := \tfrac{1}{\sqrt{2}} \left( a_3      - i a_4      \right) 
\end{align}
which will make our formulas more transparent. The energy shifts of excitations made entirely from
the creation operators $Z$ and $\phi$ are then governed by the simple effective vertices
\begin{align}
H_0  \rightarrow & \frac{M}{2}\, :\tr ( Z\bar Z+\phi\bar \phi):\\
V_{\rm eff}^{(1)} \rightarrow & - \frac{2}{M^2} : \tr \comm{Z}{\phi} \comm{\bar Z}{\bar \phi} : \label{V1}\\
V_{\rm eff}^{(2)} \rightarrow &  \frac{22N}{M^5} : \tr \comm{Z}{\phi} \comm{\bar Z}{\bar \phi} : \nonumber \\
                              & + \frac{4}{M^5} \Bigl( 
                                : \tr \comm{Z}{\phi} \comm{\bar Z}{\comm{Z}{\comm{\bar Z}{\bar \phi}}} :
+ : \tr \comm{Z}{\phi} \comm{\bar \phi}{\comm{\phi}{\comm{\bar Z}{\bar \phi}}} :
 \Bigr)\; .
\label{V2}
\end{align}
Let us stress that no limit has been performed yet, eq.~\eqref{V2} follows directly from
\eqref{dasletzte}, upon restricting its action to states given by excitations in gauge-invariant words
made of $Z$'s and $\phi$'s only. 

This result may be directly compared to the two-loop structure of the $D=4$, $\Ncal=4$ Super Yang-Mills 
dilatation operator in the analog scalar sector, which has been computed in \cite{Beisert:2003tq} eq. (5.5)  
\begin{align}
  D_0 &=  :\tr (Z\check Z +\phi\check\phi):\\
  D_2 & = \ft{\gym^2}{16\pi^2}\left (
-2 : \tr \comm{Z}{\phi} \comm{\check Z}{\check \phi} :\right ) \label{DD2}\\
  D_4 & = \ft{\gym^4}{(16\pi^2)^2}\Bigl (
4N : \tr \comm{Z}{\phi} \comm{\check Z}{\check \phi} :\nn\\&\qquad
          + 2 : \tr \comm{Z}{\phi} \comm{\check Z}{\comm{Z}{\comm{\check Z}{\check \phi}}} :
          + 2 : \tr \comm{Z}{\phi} \comm{\check \phi}{\comm{\phi}{\comm{\check Z}{\check \phi}}} : \Bigr )\label{D4}
\end{align}
The agreement of the one-loop terms was already observed for general $\grSO(6)$ excitations in the last
subsection. However,
the closeness of the two-loop result $D_4$ to $V_{\rm eff}^{(2)}$ is striking: The same structure
of terms appears, only one relative factor is wrong! Remarkably, not all terms of \eqref{D4} 
are indeed relevant in the BMN limit \eqref{BMNlimit}, as shown in \cite{Beisert:2003tq}.
However, it is precisely  the last term in \eqref{D4} which is irrelevant in the
BMN limit when acting on states (operators) of the form $\tr (Z^p\phi Z^{J-p}\phi)$ in the BMN
limit \eqref{BMNlimit}. 

So even if one  considers the BMN limit \eqref{BMNlimit} of the 
next-to-leading order effective vertex of the plane-wave matrix model
\eqref{V2}, the two-loop discrepancy to the scaling dimensions of the
four-dimensional gauge theory persists.

The observed two-loop discrepancy of plane-wave matrix theory and 
$\Ncal =4$ Super Yang-Mills should be understandable 
in the framework of Wilsonian effective quantum field theory,
where one integrates out all non-$\grSU(2)_R$-singlets in a perturbative
fashion. The resulting effective action of the $SU(2)_R$ singlet modes 
will start out with the plane-wave matrix model, but the inclusion
of the non-$SU(2)_R$-singlet modes in loops will lead
to a renormalization of the matrix model mass parameter $M$ 
as well as higher order interaction terms in the
plane-wave matrix model. As is evident from
our results at one-loop order these corrections are not yet effective, 
but apparently start to contribute beyond this order. 
It would be very interesting to study this in detail, in particular
what restrictions on these higher order interaction terms arise
from the underlying supersymmetry.


\section{Discussion}
In this paper we have shown how the mass deformed gauge quantum
mechanics of plane-wave matrix theory arises from the
Kaluza-Klein reduction of $\Ncal=4$, $D=4$ Super Yang-Mills theory
on $\R\times S^3$. Whereas the ordinary gauge quantum mechanics
is obtained through a trivial dimensional reduction of {\it any}
higher dimensional Yang-Mills theory with maximal supersymmetry,
the mass deformed supersymmetric quantum mechanics at hand
is intrinsically related to
four-dimensions, as the symmetry algebra $\grSU(2|4)$ suggests.

We went on to explore the relation of the spectrum of the
plane-wave matrix model to the scaling dimensions
of $\Ncal=4$, $D=4$ Super Yang-Mills operators. At the free
theory level the two are equivalent, provided one considers
Yang-Mills operators without higher space derivatives. Once interactions
are turned on one would in general expect a different 
behavior of the two quantities. However, for the special class 
of operators which are protected from perturbative corrections
the analogy persist: The two theories share the same 
1/2-BPS multiplets, whose primaries are given as
totally symmetric traceless $SO(6)$ tensors.
We have seen in this paper that the analogy also extends
to the one-loop level for pure scalar operators and their
supersymmetry descendants, but ceases to exist beyond that. 
It is natural to ask whether this leading-order agreement 
will turn out to be true for the
entire spectrum of the massive gauge quantum mechanics as well, 
and we hope to address this issue in the future.

The discussions presented in section 2 were purely classical, 
which implies that we can in fact start with any
supersymmetric field theory with classical superconformal
invariance and consider Kaluza-Klein reductions on $\R\times S^3$ to
obtain a mass-deformed supersymmetric gauge quantum mechanics. 
Some of them might also be given a M-theory interpretation. Let us
take the example of pure $\Ncal=2$ Yang-Mills theory, whose
trivial dimensional reduction gives a gauge quantum mechanics with
8 supersymmetries and $\grSO(5)$ symmetry. Put on $\R\times S^3$, the
mass deformation will induce the symmetry breaking
$\grSO(5)\rightarrow \grSO(3)\times \grSO(2)$. If we add one hypermultiplet
in the adjoint representation one recovers the plane-wave matrix model.
If one adds further hypermultiplets in the fundamental representation one
obtains the matrix model of supersymmetric M5-branes in a plane-wave
background, which are extended along four of the $\grSO(6)$ directions 
as well as two light-cone directions. The Supersymmetry of the relevant M5-brane
configurations in the plane-wave background is elucidated in
\cite{kimyee}. As an alternative avenue for generalization one can
also consider $\Ncal=1$ field theories with nontrivial target
spaces for chiral multiplets, i.e.~gauged nonlinear sigma models.
The relevant massive quantum mechanical models could give matrix
models of eleven dimensional pp-waves with curved transverse
spaces, which are analogs of ten dimensional solutions studied in
\cite{malmao, nkim}.

In this paper we made use of the specific formulation of
$\Ncal=4$, $D=4$ Super Yang-Mills on $\R\times S^3$, but it is
certainly quite desirable to carry out a systematic study of
supersymmetric field theories on curved backgrounds in various
dimensions, which do not allow parallel spinors. This has been
initiated by Blau in \cite{Blau:2000xg}, where pure $\Ncal=1$
Super Yang-Mills theories in $D=10,6,4,3$ and their trivial
dimensional reductions to intermediate lower dimensions are
considered. He finds that by adding mass and
Chern-Simons like terms the actions can be made supersymmetric with
respect to generalized Killing spinor equations on Einstein manifolds. 

Finally let us stress once more the emergence of an integrable $\grSO(6)$
spin chain structure at leading order perturbation theory of the plane-wave
matrix model, in analogy to the situation found at one-loop in $D=4$,
$\Ncal=4$ Super Yang-Mills by Minahan and Zarembo \cite{Minahan:2002ve}. 
It is certainly worthwhile to explore the origin and implications of this symmetry
further, possibly leading to the integrability of the complete model. 
One would expect this program to be simpler than in the full four-dimensional
case, that is, if it is performable at all.


\section*{Acknowledgments}
We would like to thank G. Arutyunov, N. Beisert, H. Nicolai, S. Kovacs, S. Metzger, A. Pankiewicz, M. P\"ossel,
T. Quella and M. Staudacher for interesting and helpful discussions.


\appendix


\section{Dimensional split of Clifford algebras} \label{cliffordsplit}

In this appendix we give the various forms of the fermionic part of the action of maximally supersymmetric
Yang-Mills theory. In 1+9 dimensions we use $G^M$-matrices\footnote{Throughout this appendix all indices are understood 
as frame indices.} satisfying the Clifford algebra $\acomm{G^M}{G^N} = 2 \eta^{MN} \unit_{32}$ with $\eta^{MN} = 
\diag(-,+^9)$. The charge conjugation matrix $C_{10}$ is symmetric and acts as $C_{10} G^M C_{10}^{-1} = {G^M}^\top$. 
In ten dimensions the fermionic field content of $\Ncal=1$ Super Yang-Mills theory is a single 32-component complex spinor 
$\Lambda_\alpha$ subject to the Majorana condition $\bar\Lambda := \Lambda^\dag G^0 \stackrel{!}{=} \Lambda^\top 
C_{10}$ as well as the Weyl condition $G_{11} \Lambda \stackrel{!}{=} \Lambda$, where $G_{11} := G^0 \cdots G^9$ is the 
chirality matrix. With these definitions the fermionic part of the $D=4$, $\Ncal=4$ theory can be written concisely as
\be
\mathcal{L}_\textindex{ferm} = \tfrac{i}{2} \bar\Lambda G^\mu D_\mu \Lambda
                             + \tfrac{1}{2} \bar\Lambda G^i \comm{\phi_i}{\Lambda} \; ,
\ee
where $\mu = 0,1,2,3$ and $i=4,\ldots,9$. We now split the $G^M$-matrices into $\Gamma^I$-matrices of $\grSO(9)$ and a 
$2\times2$ matrix factor according to
\begin{align}
G^0  & = i\sigma^2 \otimes \unit_{16}  \\
G^I  & =  \sigma^1 \otimes \Gamma^I    \qquad \mbox{with $I=1,\ldots,9$}\; .
\end{align}
The charge conjugation matrix decomposes as $C_{10} = \sigma^1 \otimes C_9$ and leads to a symmetric charge conjugation 
matrix in nine dimensions which satisfies $C_9 \Gamma^I C_9^{-1} = {\Gamma^I}^\top$.
Due to the Majorana-Weyl condition the spinor can be written as
\be
  \Lambda = \sqrt{2} {L \choose 0}  \qquad \mbox{with} \quad L^\dag = L^\top C_9
\ee
and the lagrangian takes the form
\be
\mathcal{L}_\textindex{ferm} = - i L^\dag D_t L
                               + i L^\dag \Gamma^a D_a L
                               + L^\dag \Gamma^i \comm{\phi_i}{L} \; ,
\ee
where $a=1,2,3$. We further split the $\Gamma^I$-matrices to account for $\grSO(3)\otimes\grSO(6)$:
\begin{align} \label{eqn:SO9gammas}
\Gamma^a & =
  \left( \begin{matrix}
  -\sigma^a \otimes \unit_4  &  0 \\
  0                             &  \sigma^a \otimes \unit_4
  \end{matrix} \right) &
\Gamma^i & =
  \left( \begin{matrix}
  0                              &  \unit_2 \otimes \rho^i \\
  \unit_2 \otimes {\rho^i}^\dag  &  0
  \end{matrix} \right)\; ,
\end{align}
where $\sigma^a$ are the three Pauli matrices and the $4\times4$ matrices $\rho^i$ satisfy
\be
  \rho^i {\rho^j}^\dag + \rho^j {\rho^i}^\dag = {\rho^i}^\dag \rho^j + {\rho^j}^\dag \rho^i = 2 \delta^{ij} \unit_4 \; 
.
\ee
The charge conjugation matrix in this representation is given by
\be
C_9 =
  \left( \begin{matrix}
  0                          &  -i\sigma^2 \otimes \unit_4 \\
  i\sigma^2 \otimes \unit_4  &  0
  \end{matrix} \right) \; ,
\ee
allowing one to write the spinor as
\be
L = {\lambda^{\alpha A} \choose i(\sigma^2)^{\alpha\beta} \lambda_{\beta A}^*}
\; , \qquad \alpha=1,2\:,\;\; A=1,\ldots,4 \; ,
\ee
where $\lambda^{\alpha A}$ are now four 2-component Weyl spinors. In this notation the lagrangian reads
\be
\mathcal{L}_\textindex{ferm} = - 2i \lambda_A^\dag D_t \lambda^A
                               - 2i \lambda_A^\dag \sigma^a D_a \lambda^A
                               + \lambda_A^\dag i\sigma^2 (\rho_i)^{AB} \comm{\phi_i}{\lambda_B^*}
                               - (\lambda^A)^\top i\sigma^2 (\rho_i^\dag)_{AB} \comm{\phi_i}{\lambda^B} \; .
\ee
If one finally adds the unit matrix to the set of Pauli matrices by the definition $\sigma^\mu = (\unit,\sigma^a)$, 
this is exactly the form \eqref{4daction} given in the main text.

When we reduce the field theory to the zero mode excitation in section \ref{mm_derivation}
we end up with the plane-wave matrix model where the fermions were written in $\grSU(2)\otimes\grSU(4)$ notation. 
This version of the matrix model can easily be obtained from the original form \eqref{MMlagrangian} when the above 
realization of $\Gamma$-matrices \eqref{eqn:SO9gammas} is used and the matrix model fermions are written in terms of 
Weyl spinors
\be
  \theta = {\theta^{\alpha A} \choose i(\sigma^2)^{\alpha\beta} \theta_{\beta A}^*} \; .
\ee
Note that the charge conjugation matrix $C_9$ is no longer unity, which has been used in \eqref{MMlagrangian}. We 
particularly note that
\be
\Gamma^{123} =
\left( \begin{matrix}
-i \unit_2 \otimes \unit_4  &  0 \\
0                           &  i \unit_2 \otimes \unit_4
\end{matrix} \right)
\ee
which simplifies the fermionic mass term to
\be
  \tfrac{m}{4} i \theta^\dag \Gamma_{123} \theta = \tfrac{m}{2} \theta_A^\dag \theta^A \; .
\ee


\section{Killing spinors and Killing vectors} \label{killing}

The Killing spinor equation is given by
\be \label{eqn:ksequation}
\nabla_a S^{\halpha\pm} = \pm \tfrac{i}{2R} \sigma_a S^{\halpha\pm} \; .
\ee
The hatted index $\halpha=1,2$ represents the degeneracy of the solution. When the different signs are also taken into 
account the four solutions give the lowest spinor spherical harmonics as ${\bf (2,1)\oplus (1,2)}$ of $\grSU(2)_L 
\otimes\grSU(2)_R$. In the following we will often suppress the $\pm$ index, when the given equations hold for both 
signs. The solutions of \eqref{eqn:ksequation} can be orthonormalized to
\be \label{eqn:ksorthonormal}
  S^{\halpha\dag} S^\hbeta = \delta^{\halpha\hbeta}
\ee
and form a complete basis
\be \label{eqn:kscomplete}
  \sum_\halpha S_\alpha^{\halpha} S_\beta^{\halpha*} = \delta_{\alpha\beta} \; .
\ee
Moreover they satisfy
\be \label{eqn:kssigma2}
  S^{\halpha\top} i\sigma^2 S^\hbeta = k (i\sigma^2)^{\halpha\hbeta} \quad \mbox{with} \quad k\in\C,\; \modulus{k} = 1
\ee
and we are free to define them such that $k=1$.

The Killing vectors can be obtained as bilinears $S^{\halpha\pm\dag} \sigma_a S^{\hbeta\pm}$ of (commuting)
Killing spinors. 
One can show that these bilinears, considered as $2\times2$ matrices in the indices $\halpha$ and $\hbeta$ are 
hermitian and traceless. Therefore they may be expanded into Pauli matrices:
\be
  S^{\halpha\pm\dag} \sigma_a S^{\hbeta\pm} = (\sigma_\haa)^{\halpha\hbeta} V^{\haa\pm}_a \; ,
\ee
where $\haa=1,2,3$ is a flat index. The Killing spinor equation \eqref{eqn:ksequation} for $S^\halpha$ implies the 
Killing vector equation
\be \label{eqn:kvequation}
  \nabla_a V_b^{\haa\pm} + \nabla_b V_a^{\haa\pm} = 0
\ee
for the coefficients of this expansion. Hence $V^{\haa\pm}$ give the six Killing vectors $\bf (3,1)\oplus(1,3)$ which 
are also the lowest vector spherical harmonics of $S^3$. Their orthonormality
\be \label{eqn:kvorthonormal}
  V^{\haa a} V_a^\hbb = \delta^{\haa\hbb}
\ee
can be shown from the orthonormality of the Killing spinors by making use of the Fierz identity. As an immediate 
consequence of \eqref{eqn:kvequation} we have the divergencelessness of the Killing vectors
\be \label{eqn:kvdivergenceless}
  \nabla^a V_a^\haa = 0 \; .
\ee
It can be also shown that
\be
  \nabla_a V^{\haa\pm}_b = \pm \tfrac{1}{2R} \eps^{\haa\hbb\hcc} V^{\hbb\pm}_a V^{\hcc\pm}_b
\ee
as well as
\be
  \nabla_a V^{\haa\pm}_b = \pm \tfrac{1}{R} \eps_{abc} V^{a\haa\pm} \; .
\ee

Let us also give an explicit realization of the Killing spinors and vectors. However, we emphasize that the derivation 
of the matrix model can be entirely performed without making use of them. With the obvious choice of vielbein
\be
e^1 = d\theta , \quad e^2  = \sin\theta d\psi , \quad e^3 = \sin\theta \sin\psi d\chi ,
\ee
it is a simple matter to find the solutions
\be
\label{ks}
S^{\pm} = e^{\pm i\frac{\theta}{2}\sigma^1}
          e^{i\frac{\psi}{2}\sigma^3}
          e^{i\frac{\chi}{2}\sigma^1} S_0^{\pm}
\ee
and
\begin{align}
V^{\hat{1}}_a d\x^a & = \cos\psi e^1
                        -\cos\theta \sin\psi e^2
                        \pm\sin\theta \sin\psi e^3 \; ,\nn \\
V^{\hat{2}}_a d\x^a & = \sin\psi \cos\chi e^1
                        + (\cos\theta \cos\psi \cos\chi \mp \sin\theta \sin\chi) e^2 \nn \\
                 &\quad - (\cos\theta \sin\chi \pm \sin\theta \cos\psi \cos\chi) e^3 \; ,\\
V^{\hat{3}}_a d\x^a & = \sin\psi \sin\chi e^1
                        + (\cos\theta \cos\psi \sin\chi \pm \sin\theta \cos\chi) e^2 \nn \\
                 &\quad + (\cos\theta \cos\chi \mp \sin\theta \cos\psi \sin\chi) e^3 \nn\; .
\end{align}


\section{Reduction of equation of motion} \label{derivation_eom}

In this appendix we give some details of the derivation of the equations \eqref{eqn:mm_eom} for the zero modes from the 
4-dimensional equations of motion \eqref{eqn:4d_eom}.

The first equation \eqref{eqn:eom_vector} for $\mu = 0$ can be written as
\be
  D_t(\nabla_a A^a) - i \comm{A_a}{D_t A^a} - i \comm{\phi_i}{D_t \phi_i} + 2 \acomm{\lambda^\dag}{\lambda} = 0 \; .
\ee
When the ansatz \eqref{eqn:ansatz_zero_modes} is inserted one immediately finds \eqref{eqn:mm_gauge_condition} using 
the orthogonality of Killing spinors \eqref{eqn:ksorthonormal} and Killing vectors \eqref{eqn:kvorthonormal} as well as 
the divergencelessness of the Killing vectors \eqref{eqn:kvdivergenceless}. 

For the spatial components of the same equation \eqref{eqn:eom_vector} one finds
\be
\begin{split}
& D_t^2 A_a
  - \nabla^2 A_a
  + \Rcal_a{}^b A_b
  + D_a (\nabla_b A^b)
  + i \comm{ A^b }{ 2 \nabla_b A_a - \nabla_a A_b} \\
& - \comm{A^b}{\comm{A_a}{A_b}}
  - i \comm{\phi_i}{\nabla_a \phi_i}
  - \comm{\phi_i}{\comm{A_a}{\phi_i}}
  - 2 \acomm{\lambda^\dag}{\sigma_a \lambda} = 0 \; ,
\end{split}
\ee
where $\Rcal_{ab} = \frac{2}{R^2} g_{ab}$ is the Ricci tensor of $S^3$. After inserting \eqref{eqn:ansatz_zero_modes} 
we project onto the Killing vectors. Then we use the identities
\begin{itemize}
\item
$ V^{\haa a} V^{\hbb b} \nabla_a V^{\hcc}_b
  = \frac{1}{R} \eps_{abc} V^{\haa a} V^{\hbb b} V^{\hcc c}
  = \frac{1}{R} \eps^{\haa\hbb\hcc} \det( V )
  = \frac{1}{R} \eps^{\haa\hbb\hcc} $
\item
$ S^{\halpha\dag} \sigma_a S^\hbeta V^{\haa a}
  = (\sigma_\hbb)^{\halpha\hbeta} V_a^\hbb V^{\haa a}
  = (\sigma^\haa)^{\halpha\hbeta} $
\item
$ \nabla^2 V_a^\haa = -\frac{2}{R^2} V_a^\haa $
\end{itemize}
and find \eqref{eqn:mm_eom_so3_scalars}.

The equation of motion for the scalar field \eqref{eqn:eom_scalar} becomes
\be
\begin{split}
& D_t^2 \phi_i
  - \nabla^2 \phi_i
  + \tfrac{1}{R^2} \phi_i
  - i \comm{\phi_i }{\nabla_a A^a}
  + 2i \comm{ A^a }{ \nabla_a \phi_i}
  - \comm{A^a}{\comm{\phi_i}{A_a}} \\
& - \comm{\phi_j}{\comm{\phi_i}{\phi_j}}
  + \acomm{\lambda^\dag}{i\sigma^2 \rho_i \lambda^*}
  - \acomm{\lambda^\top}{i\sigma^2 \rho_i^\dag \lambda} = 0
\end{split}
\ee
Here one has to make use of property \eqref{eqn:kssigma2}, then \eqref{eqn:mm_eom_so6_scalars} immediately follows.

The fermion equation of motion \eqref{eqn:eom_spinor} is split up into
\be
  i D_t \lambda
  - i \sigma^a \nabla_a \lambda
  + \sigma^a \comm{ A_a }{ \lambda }
  - i\sigma^2 \rho_i \comm{ \phi_i }{ \lambda^* } = 0
\ee
and projected onto the Killing spinors. Using similar identities as above one obtains \eqref{eqn:mm_eom_spinors}.


\section{A useful identity for gauge invariant states} \label{sec:misc}

The operator $: \tr T^a \comm{a_i^\dag}{a_i} :$ annihilates any $n$-trace state
\be
 \tr (a_{j_{1,1}}^\dag \cdots a_{j_{1,k_1}}^\dag) \tr (a_{j_{2,1}}^\dag \cdots a_{j_{2,k_2}}^\dag) \cdots \tr 
(a_{j_{n,1}}^\dag \cdots a_{j_{n,k_n}}^\dag) \ket{0} \; .
\ee
This follows from the facts that it commutes with a trace of creation operators and annihilates the vacuum. It commutes 
since the sum of all Wick contractions vanishes
\be\nn
\begin{split}
  : \tr T^a \comm{a_i^\dag}{\rnode{AAA}{a_i}} : \tr (a_{j_1}^\dag \rnode{BBB}{\cdots} a_{j_k}^\dag)
& = \ : \tr \comm{T^a}{a_i^\dag} \rnode{CCC}{a_i} : \tr (a_{j_1}^\dag \rnode{DDD}{\cdots} a_{j_k}^\dag) \\[3mm]
& = \tr (\comm{T^a}{a_{j_1}^\dag} a_{j_2}^\dag \cdots
a_{j_k}^\dag)
  + \tr (a_{j_1}^\dag \comm{T^a}{a_{j_2}^\dag} \cdots a_{j_k}^\dag) \\
&\qquad   + \tr (a_{j_1}^\dag a_{j_2}^\dag \cdots \comm{T^a}{a_{j_k}^\dag}) \\
& = \tr (T^a a_{j_1}^\dag a_{j_2}^\dag \cdots a_{j_k}^\dag) - \tr (a_{j_1}^\dag a_{j_2}^\dag \cdots a_{j_k}^\dag T^a) = 
0 \; .
\ncbar[nodesep=3pt,arm=1.5mm,angle=-90]{AAA}{BBB}
\ncbar[nodesep=3pt,arm=1.5mm,angle=-90]{CCC}{DDD}
\end{split}
\ee


\section{Effective 2-loop vertex} \label{sec:perturbation_theory_2-loop}

Summing all 2-loop contributions, we find
\be\nn
\begin{split}
V_{\rm eff}^{(2)} \rightarrow \
& - \frac{51N^2}{4M^5} : \tr a_i^\dag a_i : \\
& - \frac{2}{M^5} \Bigl( : \tr a_i a_j \tr a_i^\dag a_j^\dag :
                       + : \tr a_i a_i^\dag \tr a_j a_j^\dag :
                       + : \tr a_i a_j^\dag \tr a_i a_j^\dag : \\
& \qquad\qquad         + : \tr a_i a_i \tr a_j^\dag a_j^\dag :
                     + 2 : \tr a_i a_j^\dag \tr a_i a_j^\dag : \Bigr) \\
& - \frac{N}{M^5} \Bigl( - 30 : \tr a_i a_j a_i^\dag a_j^\dag :
              + \tfrac{67}{8} : \tr a_i a_i^\dag a_j a_j^\dag :
              + \tfrac{67}{8} : \tr a_i a_j^\dag a_j a_i^\dag :
              + \tfrac{61}{4} : \tr a_i a_j a_j^\dag a_i^\dag : \\
& \qquad\qquad           + 17 : \tr a_i a_i a_j^\dag a_j^\dag :
                         - 15 : \tr a_i a_j^\dag a_i a_j^\dag : \Bigr) \\
& - \frac{1}{M^5} \Bigl( : \tr \comm{a_j}{\comm{a_i}{a_j^\dag}} \comm{a_k}{\comm{a_i^\dag}{a_k^\dag}} :
                       + : \tr \comm{a_j^\dag}{\comm{a_i}{a_j}} \comm{a_k}{\comm{a_i^\dag}{a_k^\dag}} : \\
& \qquad\qquad         + : \tr \comm{a_j}{\comm{a_i}{a_j^\dag}} \comm{a_k^\dag}{\comm{a_i^\dag}{a_k}} :
                       + : \tr \comm{a_j^\dag}{\comm{a_i}{a_j}} \comm{a_k^\dag}{\comm{a_i^\dag}{a_k}} : \\
& \qquad\qquad         + : \tr \comm{a_j}{\comm{a_i^\dag}{a_j^\dag}} \comm{a_k}{\comm{a_i^\dag}{a_k}} :
                       + : \tr \comm{a_j^\dag}{\comm{a_i^\dag}{a_j}} \comm{a_k}{\comm{a_i^\dag}{a_k}} : \\
& \qquad\qquad         + : \tr \comm{a_j}{\comm{a_i}{a_j^\dag}} \comm{a_k^\dag}{\comm{a_i}{a_k^\dag}} :
                       + : \tr \comm{a_j^\dag}{\comm{a_i}{a_j}} \comm{a_k^\dag}{\comm{a_i}{a_k^\dag}} : \\
& \qquad\qquad         + : \tr \comm{a_j}{\comm{a_i^\dag}{a_j}} \comm{a_k^\dag}{\comm{a_i}{a_k^\dag}} :
          - \tfrac{1}{2} : \tr \comm{a_j}{\comm{a_i}{a_j}} \comm{a_k^\dag}{\comm{a_i^\dag}{a_k^\dag}} : \Bigr)
\end{split}
\ee
This can be simplified using Jacobi identities and a reasoning similar to that at one loop, e.~g. we have
\be
\begin{split}
: \tr \comm{a_j}{\comm{a_i}{a_j^\dag}} \comm{a_k}{\comm{a_i^\dag}{a_k^\dag}} :
& = \                  : \tr \comm{a_j^\dag}{\comm{a_i}{a_j}} \comm{a_k}{\comm{a_i^\dag}{a_k^\dag}} :
         + N : \tr \comm{a_i}{a_j} \comm{a_i^\dag}{a_j^\dag} : \\
& \qquad + : \tr \comm{a_i}{\comm{a_n}{\comm{a_i^\dag}{a_n^\dag}}} T^a : \ : \tr T^a \comm{a_j^\dag}{a_j} :
\end{split}
\ee
The last line may be dropped by arguments given in appendix \ref{sec:misc}.

Then we find the somewhat more compact final result
\be\label{dasletzte}
\begin{split}
V_{\rm eff}^{(2)} \rightarrow \
& - \frac{2}{M^5} \Bigl( : \tr a_i a_i \tr a_j^\dag a_j^\dag :
                     + 2 : \tr a_i a_j^\dag \tr a_i a_j^\dag : \Bigr) \\
& - \frac{N}{M^5} \Bigl( - 11 :\tr \comm{a_i^\dag}{a_j^\dag} \comm{a_i}{a_j}: 
          + 17 : \tr a_i a_i a_j^\dag a_j^\dag :
                         - 15 : \tr a_i a_j^\dag a_i a_j^\dag : \Bigr) \\
& - \frac{4}{M^5} : \tr \comm{a_j^\dag}{\comm{a_i}{a_j}} \comm{a_k}{\comm{a_i^\dag}{a_k^\dag}} : \\
& - \frac{1}{M^5} \Bigl( : \tr \comm{a_j}{\comm{a_i^\dag}{a_j^\dag}} \comm{a_k}{\comm{a_i^\dag}{a_k}} :
                       + : \tr \comm{a_j^\dag}{\comm{a_i^\dag}{a_j}} \comm{a_k}{\comm{a_i^\dag}{a_k}} : \\
& \qquad\qquad         + : \tr \comm{a_j}{\comm{a_i}{a_j^\dag}} \comm{a_k^\dag}{\comm{a_i}{a_k^\dag}} :
                       + : \tr \comm{a_j^\dag}{\comm{a_i}{a_j}} \comm{a_k^\dag}{\comm{a_i}{a_k^\dag}} : \\
& \qquad\qquad         + : \tr \comm{a_j}{\comm{a_i^\dag}{a_j}} \comm{a_k^\dag}{\comm{a_i}{a_k^\dag}} :
          - \tfrac{1}{2} : \tr \comm{a_j}{\comm{a_i}{a_j}} \comm{a_k^\dag}{\comm{a_i^\dag}{a_k^\dag}} : \Bigr)
\end{split}
\ee


\end{document}